\documentclass[aps,floatfix,prb,reprint,superscriptaddress]{revtex4-1}
\usepackage{graphicx}
\usepackage{units}
\newcommand{\vecbf}[1]{\mbox{\boldmath$#1$}}
\begin{document}
\title{Superconducting coplanar waveguide resonators for low temperature pulsed electron spin resonance spectroscopy}
\author{H. Malissa}
\affiliation{Department of Electrical Engineering, Princeton University}
\author{D. I. Schuster}
\affiliation{Department of Physics, University of Chicago}
\author{A. M. Tyryshkin}
\affiliation{Department of Electrical Engineering, Princeton University}
\author{A. A. Houck}
\affiliation{Department of Electrical Engineering, Princeton University}
\author{S. A. Lyon}
\affiliation{Department of Electrical Engineering, Princeton University}
\begin{abstract}
We discuss the design and implementation of thin film superconducting coplanar waveguide micro-resonators for pulsed ESR experiments. The performance of the resonators with P doped Si epilayer samples is compared to waveguide resonators under equivalent conditions. The high achievable filling factor even for small sized samples and the relatively high Q-factor result in a sensitivity that is superior to that of conventional waveguide resonators, in particular to spins close to the sample surface. The peak microwave power is on the order of a few microwatts, which is compatible with measurements at ultra low temperatures. We also discuss the effect of the nonuniform microwave magnetic field on the Hahn echo power dependence.
\end{abstract}
\maketitle
Planar micro-resonators have been demonstrated to be an attractive alternative to conventional waveguide resonators for electron spin resonance (ESR) spectroscopy in situations where the sample volume and therefore the number of spins are small\cite{Johansson1974,Wallace1991,Mamin2003,Narkowicz2005,Narkowicz2008,Twig2011,Torrezan2009}. The frequency of micro-resonators is determined only by their length along the direction of propagation, whereas both perpendicular dimensions are much shorter than the wavelength. The resonator volume can thus be tailored to be close to the sample volume. Here we report the design and fabrication of superconducting CPW resonators for pulsed ESR measurements. The superconducting metallization layer is made to be thin enough to support the static magnetic field required of ESR at \unit[10]{GHz}. The performance of the devices is demonstrated with spin echo measurements on an isotopically purified $^{28}$Si:P.

In contrast to waveguide resonators which are optimized to exhibit a very uniform microwave magnetic field $B_{1}$ across the sample volume, the mode pattern of planar micro-resonators is determined by the geometry of the resonator and the planar waveguide. $B_{1}$ can therefore be quite inhomogeneous across the sample volume, which make the resonators particularly sensitive to spins close to the surface. The high conversion efficiency of microwave power to $B_{1}$ makes these resonators convenient for low temperature operation where the cooling power is limited and microwave heating must be avoided.

In coplanar waveguides (CPW) the center conductor and the ground planes are on the same side of the substrate and are separated by a gap. The impedance is mostly independent of the substrate thickness, thus allowing the waveguide cross section to be arbitrarily small, which further reduces the resonator volume. The use of thin film superconductors as center conductor and ground plane metallization layer supports the resulting high current densities and allows for much higher quality factors. CPW resonators are used in connection with superconducting qubits and continuous wave ESR has been demonstrated. In particular, a hybrid system consisting of a superconducting qubit and a spin system that serves as a memory is feasible\cite{Wesenberg2009,Schuster2010,Kubo2010,Kubo2011,Zhu2011}. 

\begin{figure}
\includegraphics[width=\columnwidth]{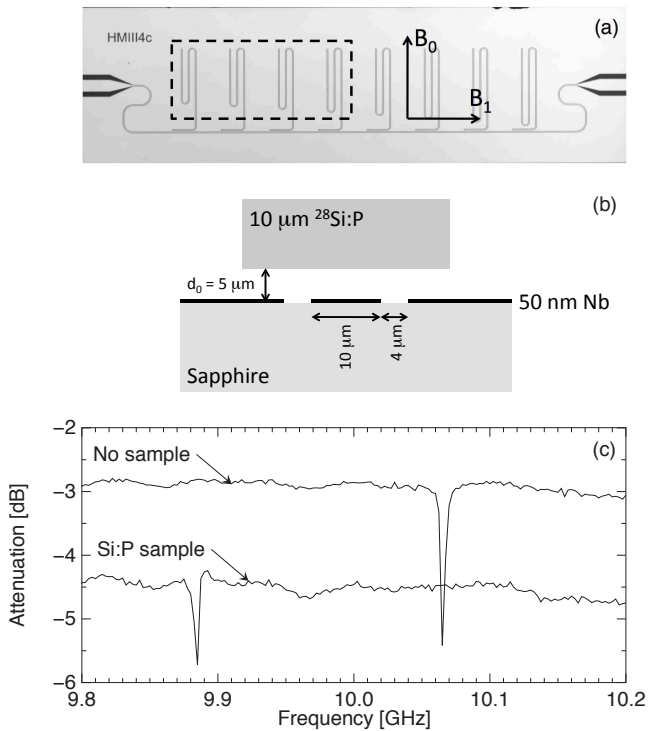}
\caption{(a) Photograph of the device, showing eight $\lambda/4$ resonators coupled to the common feedline. The Nb film appears gray, and the underlying sapphire substrate is black. The whole device is $\unit[7\times 2]{mm}$ in size. The dashed rectangle indicates the position where silicon sample was mounted. The arrows indicate the orientation of the external field $B_{0}$ and the oscillating field $B_{1}$. (b) Cross section of the CPW resonator and the active sample volume. (c) Resonance curves of one of the CPW resonators before and after mounting a silicon sample, measured using a vector network analyzer. The resonance width in both cases is approximately \unit[5]{MHz} which corresponds to $Q\approx 2000$.}
\end{figure}

The superconducting CPW structures are fabricated in \unit[50]{nm} Nb films sputtered on sapphire substrates. The thickness of \unit[50]{nm} is selected to ensure that the CPW remains superconducting in the magnetic fields used for X-band ESR experiments (typically \unit[350]{mT} at \unit[10]{GHz}). The theoretical in-plane critical field of these films\cite{Wallace1991} is \unit[1.5]{T} at \unit[4.2]{K}, compared to \unit[200]{mT} for bulk Nb. The resonators and waveguides are patterned using photolithography followed by plasma etching. The devices, with a size of $\unit[7\times 2]{mm}$, consist of eight $\lambda/4$ sections of CPW which are capacitively coupled to a common feedline on one end and shorted to ground on the other end (Fig.~1(a)). The coupling is achieved by a short L-shaped section of the resonator running in parallel to the feedline. The length of the individual resonators varies from \unit[2.8]{mm} to \unit[3.4]{mm}, which corresponds to microwave frequencies ranging from \unit[8.75]{GHz} to \unit[10.5]{GHz} for the given waveguide geometry and the relative dielectric constant of sapphire ($\epsilon_{r}=11.5$). The CPW has a center conductor width of $\unit[10]{\mu m}$ and a gap width of $\unit[4]{\mu m}$, much smaller than the wavelength. Near the edge of the chip the waveguide is tapered to a width of $\unit[150]{\mu m}$ to facilitate interconnection through wire bonding. 

The sample examined is a $\unit[\sim 3\times 1]{mm^{2}}$ piece of isotopically enriched $^{28}$Si:P epilayer of $\unit[10]{\mu m}$ thickness with a doping concentration of $\unit[1\times 10^{16}]{cm^{-3}}$ (indicated as rectangle in Fig.\ 1(a)). This sample has been investigated intensively in the past, and the decoherence time of $T_{2}=\unit[280]{\mu s}$ and an ESR linewidth of $\unit[10]{\mu T}$ were reported\cite{Tyryshkin2003} for P donors at \unit[7]{K}. The signal is comparably strong, which makes this sample ideal for demonstration and evaluation purposes. The sample is flip-chip mounted directly on top of the resonator (see Fig.~1(b)), and the device is oriented such that the external field $B_{0}$ is in plane of the device and perpendicular to the microwave $B_{1}$ field (see Fig.~1(a)).

A vector network analyzer (Anritsu MS2028C) is used to characterize the resonators at \unit[4.2]{K} before and after the sample is mounted (traces are shown in Fig.~1(c)). The resonances typically have a width of \unit[5]{MHz}, and their Q-factor is thus $Q\approx 2000$. The dielectric constant of Si ($\epsilon_{r}=11.9$) is high, explaining a significant shift in the resonant frequency upon mounting the silicon sample. The Q-factor is limited by the temperature and is expected to increase significantly below \unit[4.2]{K}.

The exact orientation of the static magnetic field with respect to the thin film superconductor is critical, since a perpendicular field component of approximately \unit[1]{mT} is sufficient to be threaded through the film. Trapped flux remains in the film, causing local field distortions and significant hysteretic frequency shifts of the resonance and a decrease in resonator Q-factor\cite{Wallace1991}. We reproducibly achieved satisfactory alignment by gradually increasing the external field from zero to its target value of \unit[350]{mT} while compensating any frequency shifts and Q-factor reductions by carefully readjusting the orientation of the superconducting film. Whenever the misorientation was too large and flux was trapped, the superconductor was annealed by raising its temperature above the transition temperature. High-Q resonances have been observed up to fields of \unit[1]{T}.

\begin{figure}
\includegraphics[width=\columnwidth]{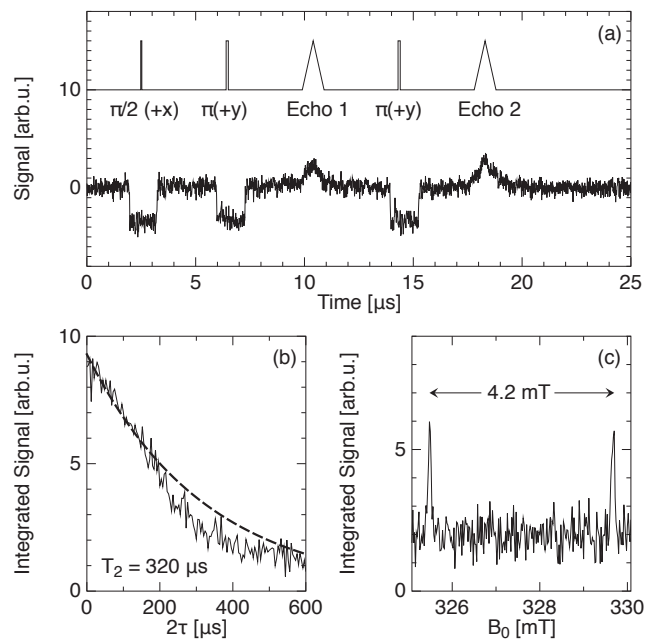}
\caption{(a) Hahn echo signals (bottom trace) measured at \unit[4.2]{K} with the pulse sequence shown at top. Two refocussed echoes are seen after two refocussing pulses. The $\pi/2$ and $\pi$ pulses are \unit[50]{ns} and \unit[100]{ns} long, respectively. The pulses in the experimental trace appear broader than they are because of overlapping defense pulses. (b) Hahn echo decay as a function of $2\tau$ (solid line), along with an exponential fit with time constant $T_{2}=\unit[320]{\mu s}$ (dashed line). The decay curve does not follow the exponential for $2\tau>\unit[200]{\mu s}$ due to field fluctuations in the magnet (see Ref.~\onlinecite{Tyryshkin2006}). (c) An ESR spectrum measured by integrating the echo signal while sweeping the magnetic field. The two hyperfine split lines are $\unit[4.2]{mT}$ apart as expected for P donors.}
\end{figure}

Pulsed ESR experiments are performed using a commercial X-band spectrometer (Bruker ElexSys 580). A Hahn echo sequence with two refocussing pulses [$\pi/2(+x)$ -- $\tau$ -- $\pi(+y)$ -- $\tau$ -- echo1 -- $\tau$ -- $\pi(+y)$ -- $\tau$ -- echo2] is employed to measure $T_{2}$ of P donors. $T_{1}$ of donors is longer than \unit[10]{s} at \unit[4.2]{K} (Ref.~\onlinecite{Feher1959}). In order to allow faster acquisition time not limited by $T_{1}$, we used an LED pre-flash of \unit[100]{ms} followed by a delay of \unit[50]{ms} to thermalize the spins before running the echo pulse sequence. Fig.~2(a) illustrates a typical time-domain transient with two echo signals after 1000 signal averaging accumulations. The echo amplitude decays exponentially with $2\tau$ and is a direct measure of $T_{2}$ (Fig.~2(b)). From the fit we establish $T_{2}=\unit[320]{\mu s}$, in excellent agreement with previous measurements.\cite{Tyryshkin2003} In the field sweep experiment in Fig.~2(c) the delay time $2\tau$ is kept constant, whereas the external field is scanned. The two hyperfine lines split by \unit[4.2]{mT} are resolved providing a definite signature of P donors. 
The echoes shown in Fig.~2(a) have a FWHM of approximately $\unit[1]{\mu s}$. This corresponds to a linewidth of \unit[0.01]{mT}, close to the value established for this sample in previous measurements. This means that the superconducting layer does not distort the static magnetic field substantially. The microwave pulse power used in these experiments was $P=\unit[2]{mW}$, four orders of magnitude lower than for a waveguide resonator in an equivalent experiment. 

\begin{figure}
\includegraphics[width=\columnwidth]{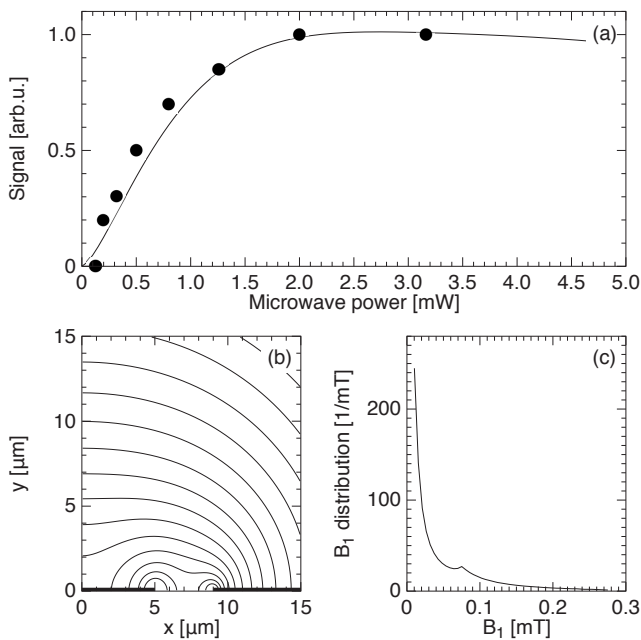}
\caption{(a) Microwave power dependence of the Hahn echo signal intensity (measured data: solid circles; simulation: line). (b) A contour plot of microwave magnetic field magnitude $B_{1}$ in one quadrant in the plane perpendicular to the CPW. The contour lines are spaced logarithmically with the field decreasing by 20\% from one contour to the next. The location of center conductor and ground plane metallization are indicated as bold black lines.  (c) Distribution of calculated $B_{1}$ values at applied microwave power of \unit[2]{mW}.}
\end{figure}

The distribution of $B_{1}$ fields in the plane perpendicular to the CPW is determined by the characteristic geometry of the CPW signal and ground plane lines, the dielectric constant above and below the waveguide, and the sample thickness. Fig.~3(b) shows the calculated $B_{1}$ magnitude for our particular geometry. $B_{1}$ is inhomogeneous across the sample volume, reaching its maximum in the gap between signal and ground lines and decreasing rapidly towards the sample interior. 

The rotation angle $\theta$ of a rectangular pulse of strength $B_{1}\propto\sqrt{P}$ and duration $t_{p}$ is given by
\[\theta=g\mu_{B}B_{1}t_{p}/\hbar\] 
where $g$ is the electron g-factor and $\mu_{B}$ is the Bohr magneton. Because of the broad $B_{1}$ distribution only a fraction of the spins experience the desired rotation $\theta$, whereas all other spins are either over or under rotated. In waveguide resonators that have a narrow $B_{1}$ distribution (e.g. less than 10\%), the echo amplitude after applying the Hahn echo sequence [$\theta_{1}$ -- $\tau$ -- $\theta_{2}$ -- $\tau$ -- echo] with $\theta_{2}=2\theta_{1}$ varies as $\sin^{3}\theta_{1}$, exhibiting pronounced oscillations of the echo amplitude as a function of $B_{1}$ with a maximum at $\theta_{1}=\pi/2$ and $\theta_{2}=\pi$. In comparison, because of the broad $B_{1}$ distribution such oscillations are not seen in the power dependence measured in our CPW resonators (Fig.~3(a)). Instead, the dependence is barely peaked, reaching the maximum intensity at microwave power around \unit[2]{mW} and remaining flat at higher powers. 

We simulated the power dependence in Fig.~3(a) by summing up the contributions of all spins and taking into account the field strength $B_{1}(\vecbf{r})$ at each spin location \vecbf{r}:\cite{Hoult2001}
\begin{eqnarray*}
\mbox{signal}\left(P\right)&=&N\left<\sin^{3}\theta\left(P,\vecbf{r}\right)\hbar g_{S}\left(\vecbf{r}\right)\right>_{\vecbf{r}}\\ 
&\propto&\left<\sin^{3}\theta\left(P,\vecbf{r}\right)B'_{1}\left(\vecbf{r}\right)\right>_{\vecbf{r}}
\end{eqnarray*}
where $g_{S}(\vecbf{r})$ is the spin-resonator coupling constant\cite{Schuster2010,Kubo2010} and $N$ is the total number of spins. $B'_{1}\left(\vecbf{r}\right)$ is the microwave magnetic field normalized to unit power and is determined only by the CPW parameters. $B'_{1}$ is proportional to $g_{S}$, since $g_{S}$ is the vacuum Rabi frequency arising from the microwave magnetic field at the spin location $\vecbf{r}$ in the ground state of the resonator. Our model thus has only three fitting parameters: the distance $d_{0}$ of the sample to the CPW metallization layer (Fig.~1(b)), the microwave power to magnetic field conversion efficiency, and an amplitude scaling factor. All three can be established by fitting the model to the measured data (solid line in Fig.~3(a)). For example, the extracted $d_{0}$ is $\unit[5]{\mu m}$, which is quite reasonable. 

The extracted microwave conversion efficiency allows us to calculate the distribution of absolute $B_{1}$ fields at each applied microwave power. Fig.~3(c) shows the calculated $B_1$ field distribution at \unit[2]{mW} pulse power that corresponds to the power where the echo amplitude saturates as seen in Fig.~3(a)). The calculated distribution diverges at low $B_{1}$ fields because the number of contributing spins increases with distance from the waveguide. However, these remote spins do not contribute to the signal and our calculations show that the nearby spins within a lateral distance of $\unit[15]{\mu m}$ from the waveguide constitute 95\% of the total echo signal.

The resonator volume can thus be assumed to be $V=\unit[30]{\mu m}\times\unit[10]{\mu m}\times\unit[3]{mm}$ which gives a total number of spins of $N\approx 4.5\times 10^{9}$ per hyperfine split line for the given doping concentration. The signal-to-noise ratio established from transient measurements (Fig.~2(a)), scaled to a single shot (no signal averaging) is approximately $\sim 0.63$, and $\sim 10$ when the echo signal is integrated over its width of $\unit[1]{\mu s}$. This corresponds to a single shot sensitivity of $4.5\times 10^{8}$ spins. This number is two orders of magnitude better than the sensitivity of a waveguide resonator under comparable conditions. We estimate that the sensitivity can be further improved by a factor of 2.5 if the sample is mounted closer to the resonator surface ($d_{0}=\unit[100]{nm}$ instead of $\unit[5]{\mu m}$). The sensitivity is expected to improve even further when the resonator is cooled down below \unit[4.2]{K}, which would lead to significantly higher Q-factor.

Our spin number sensitivity compares favorably with those reported for other micro-resonator structures. \citet{Narkowicz2005} report a signal-to-noise ratio of 560 after 1000 acquisitions from \unit[$9.4\times 10^{14}$]{spins} using a lumped-element resonator with a coil diameter of \unit[0.2]{mm} and a DPPH sample. Scaled to a sensitivity per shot, this corresponds to \unit[$3\times 10^{13}$]{spins}, which is five orders lower than in our case. \citet{Twig2011} report a sensitivity of $\unit[1.5\times 10^{6}]{spins/\sqrt{Hz}}$ from a CPMG sequence using a non-superconducting surface loop-gap resonator and a Si:P sample that is similar to ours. Their experiments are done at $T=\unit[10]{K}$ where $T_{1}\approx T_{2}$, allowing an effective acquisition rate of \unit[100]{kHz} (100 CPMG echoes at a rate of \unit[1]{kHz}). The sensitivity per shot is thus $4.8\times 10^{8}$, quite comparable to our results.

In conclusion, we have fabricated a thin film superconducting coplanar waveguide resonator and evaluated its performance with pulsed ESR spectroscopy on $^{28}$Si:P epilayer samples. The small resonator volume and consequently its high filling factor leads to a high sensitivity, in particular towards spins close to the interface between resonator and sample. The peak power required for microwave pulses is in the \unit{mW} range, which is compatible with low temperature operation. The superconducting layer does not introduce a significant distortion of the external magnetic field when the resonator is oriented parallel to the field. The inhomogeneous field distribution leads to a complex dependence of echo intensity on microwave power which can be accurately described by model calculations taking the specific resonator geometry into account. This field inhomogeneity limits the fidelity of multi-pulse sequences, which should be partly resolved by using adiabatic pulses\cite{Baum1985,DeGraaf1997}. The resonator is a useful alternative to conventional waveguide resonators for studying small numbers of electron spins at or near a surface.
\bibliographystyle{apsrev4-1}
\bibliography{../../../JabRef/Alexia}
\end{document}